\documentclass[amsmath,showpacs,prb,twocolumn] {revtex4-1}
\usepackage{bm}
\usepackage{enumerate}
\usepackage{graphicx}
\usepackage[dvips]{epsfig}

\makeatletter
\newcommand{\Rmnum}[1]{\expandafter\@slowromancap\romannumeral #1@}
\makeatletter

\begin{document}
\title{Spatially anisotropic kagome antiferromagnet with Dzyaloshinskii-Moriya interaction}
\author{Vladimir A. Zyuzin}
\author{Gregory A. Fiete}
\affiliation{Department of Physics, The University of Texas at Austin, Austin, TX 78712, USA}

\begin{abstract}
We theoretically study the spatially anisotropic spin-1/2 kagome antiferromagnet with Dzyaloshinskii-Moriya (DM) interaction using a renormalization group analysis in the quasi-one-dimensional limit.  We identify the various temperature and energy scales for ordering in the system.  For very weak DM interaction, we find a low-temperature spiral phase with the plane of the spiral selected by the DM interaction. This phase is similar to a previously identified phase in the absence of the DM interaction.  However, above a critical DM interaction strength we find a transition to a phase with coexisting antiferromagnetic and dimer order, reminiscent of one-dimensional antiferromagnetic systems with a uniform DM interaction.  Our results help shed light on the fate of two dimensional systems with both strong interactions and significant spin-orbit coupling.
\end{abstract}


\maketitle

\section{Introduction}

In recent years, spin-orbit coupling has become a central theme in condensed matter physics.\cite{Nagaosa:rmp10,Zutic:rmp04}  Much attention has been focused on time-reversal invariant topological insulators (TI) which are driven by strong spin-orbit coupling in the weak interaction limit.\cite{Moore:nat10,Hasan:rmp10,Qi:rmp11}   Of central interest is the nature of the phases realized when both spin-orbit coupling and interactions are significant.\cite{Fiete:sci11}  A number of studies have investigated the fate of the TI phase in two and three dimensions as the strength of the interactions is increased,\cite{Pesin:np10,Kargarian:prb11,Witczak-Krempa:prb10,Ruegg11,Rachel:prb10,Hohenadler:prl11,Yamaji:prb11,Wu:prl11} which reveals some possible outcomes of the competition between interactions and spin-orbit coupling.  Under some conditions, quantum spin liquids\cite{Hohenadler:prl11,Ruegg11,Wu:prl11} and other exotic states are expected,\cite{Young:prb08,Pesin:np10,Kargarian:prb11} which suggests interesting connections between TI and spin systems.\cite{Fiete:pe11}

In this work, we study a highly frustrated spin model--the spin-1/2 kagome antiferromagnet with spatially anisotropic exchange constants and spin-orbit coupling.   We investigate the physics of this system as the spin-orbit coupling, realized in the form of Dzyaloshinskii-Moriya (DM) interaction,\cite{Dzyaloshinskii,Moriya} is increased. The ground state of the spatially isotropic spin-1/2 Heisenberg antiferromagnet in the absence of DM interaction has been studied for some time.\cite{Marston:jap91,Sachdev:prb92,Singh:prb07,Elser1,Elser2,Young,Lecheminant:prb97,Sindzingre:prl00,Singh:prl92,Misguich:prb05,hermele,Evenbly:prl10} Its large classical ground state degeneracy was thought to make it an excellent candidate for a quantum spin liquid, and the most recent density matrix renormalization group results seem to indicate this is indeed the case.\cite{White:sci11} Exactly solvable spin-3/2 spin liquids have also been studied on the isotropic kagome lattice.\cite{Chua:prb11}

However, some kagome antiferromagnets, such as volborthite, Cu$_{3}$V$_{2}$O$_{7}$(OH)$_{2}\cdot 2$H$_{2}$O, are spatially anisotropic which allows a quasi-one dimensional approach to be used in their study.\cite{Oleg1,Oleg2,Oleg3,Oleg4,Oleg5}  This approach is especially powerful because one-dimensional methods are  adept at describing strongly correlated physics.\cite{Giamarchi,Gogolin}  In this work, we extend earlier studies of the DM interaction on the spatially isotropic kagome lattice\cite{Rigol,Rigol:prl07,Hermele:prb08,Elhajal:prb02,Cepas1,Cepas2,Sachdev2,Tchernyshyov,kagome_dm1,kagome_dm2,herbert_dm1, herbert_dm2, herbert_dm3} to the spatially {\em anisotropic} case by using a quasi-one dimensional approach.  Our main result is that we find a transition from a spiral ordered state with weak Neel order perpendicular to the spiral\cite{Oleg4} to a spiral state with weak in-plane Neel order coexisting with dimer order as the strength of the DM interaction is tuned up. (See Fig.\ref{fig:kagome} and Fig.\ref{fig:phase_diagram}.)  We argue this transition may be in an experimentally accessible parameter range and could possibly be tuned by pressure.

Before we give a detailed account of our theoretical results, we first begin with a brief survey of the growing number experimental spin-1/2 kagome systems.

\subsection{Brief survey of related experimental systems}

A major motivation for the experimental and theoretical study of spin-1/2 kagome antiferromagnets is the possibility of spin liquid and other highly fluctuating phases.\cite{Balents_Nature,White:sci11}  The kagome lattice leads to geometrical frustration which works against an ordered state, as does the small size of the spin which enhances quantum fluctuations.  However, there is theoretical evidence that the DM interaction, allowed by symmetry on the kagome lattice, tends to drive a spin liquid towards a magnetically ordered state.\cite{Cepas1,Sachdev2}  In practice, then, it is important to have an estimate of the size of the DM interaction in a given system.  Experimental realities such as defects and lattice anisotropy may also influence the low-temperature phases.  It is thus important to include these aspects in theoretical studies as well.  In this paper, we focus on the role of lattice anisotropy in spin-1/2 kagome antiferromagnets with a DM interaction.

Among the experimentally studied spin-1/2 kagome materials, an absence of long-range magnetic order down to small temperatures (compared to the characteristic exchange energy) is common.   Important examples of spin-1/2 kagome antiferromagnets include herbertsmithite, ZnCu$_{3}$(OH)$_{6}$Cl$_{2}$,\cite{herbert1,herbert2,herbert_review,Okamoto:prb11,Olariu:prl08,Han:prb11,Helton:prl10}  vesignieite BaCu$_{3}$V$_{2}$O$_{8}$(OH)$_{2}$,\cite{vesignieite1, vesignieite2,Okamoto:prb11,Coleman:prb11} volborthite Cu$_{3}$V$_{2}$O$_{7}$(OH)$_{2}\cdot 2$H$_{2}$O,\cite{volborthite1, volborthite2, volborthite3,Yoshida:prb11}   MgCu${}_{3}$(OH)${}_{6}$Cl${}_{2}$ (a cousin of herbertsmithite),\cite{Kermarrac:prb11} and Rb$_{2}$Cu$_{3}$SnF$_{12}$.\cite{pinwheel}

\begin{figure}
\includegraphics[width=0.9\linewidth]{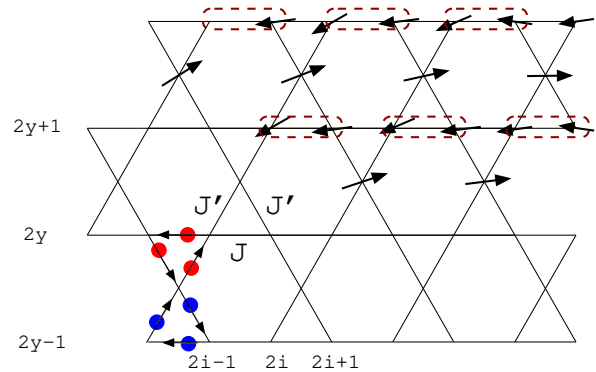}
\caption{(color online) The kagome lattice with spatially anisotropic exchange couplings.  Sites along the one-dimensional spin chains (running horizontally) are labeled by $i$ as indicated, and different horizontal chains are labeled by $y$.   The scale of the antiferromagnetic nearest neighbor spin exchange along the chains is $J$, and the scale of the corresponding exchange between spins on a chain and the "middle spins" (between chains) is $J'$ as shown.  We consider the anisotropic limit $J' \ll J$ in this paper.  Our convention for the staggered Dzyaloshinskii-Moriya (DM) interaction is indicated by the different colors on the "up" and "down" triangles of the kagome lattice, with the up triangles having a DM vector pointing out of the kagome plane and a down triangle having a DM vector pointing into the plane of the kagome.  The arrows indicated the direction going from site $i\to j$.  The strength of the DM interaction is given by \eqref{eq:DM}, and shares the same anisotropy as $J$ and $J'$ ({\it i.e.} $D^{z'} \ll D^z$) due to the spatial anisotropy of the lattice. Black arrows and brown dashed rectangles correspond to a DM interaction driven spin ordering we found in this paper. Arrows stand for the spin direction while brown dashed rectangles describe dimers.  Shown is the ``Neel+dimer" order given in Eq.\eqref{result1}.}\label{fig:kagome}
\end{figure}

Herbertsmithite has isotropic exchange interactions and has received perhaps the largest amount of theoretical and experimental attention among the materials listed above. No magnetic order appears in this system down to temperatures of 40 mK (nearest-neighbor exchange $J \sim 180$K) making it a leading quantum spin liquid candidate.\cite{herbert1, herbert2, herbert_review,Okamoto:prb11,Olariu:prl08,Han:prb11,Helton:prl10} Vesignieite also possesses an isotropic kagome lattice and exhibits similar behavior to herbertsmithite over a wide temperature range ($J \sim 53$K).\cite{vesignieite1, vesignieite2,Coleman:prb11} However, it was recently observed that vesignieite exhibits an exotic phase at very low temperatures which appears to be described by the coexistence of dynamical and small frozen moments.\cite{Coleman:prb11} The DM interaction, with an energy scale set by $D$, may play an important role in understanding the low temperature phase in  vesignieite ($D/J \approx 0.14$).\cite{Coleman:prb11} Volborthite is characterized by a small spatial anisotropy in the exchange interactions between spins, and no magnetic order has been reported down to temperatures of order $2K$ (with $J\sim 77$K).\cite{volborthite1, volborthite2, volborthite3,Yoshida:prb11} However, a signature of frozen magnetic moments appears below this temperature scale.\cite{volborthite1, volborthite2, volborthite3,Yoshida:prb11}  The low-temperature magnetic order may be a consequence of spatial anisotropy in the exchange interactions.\cite{Oleg4,balents1}  One of the main advantages in studying volborthite over herbertsmithite and vesignieite is that it can be prepared with fewer impurities.\cite{volborthite1, volborthite2, volborthite3,Yoshida:prb11}  In another deformed kagome compound, Rb$_{2}$Cu$_{3}$SnF$_{12}$,\cite{pinwheel}  a valence bond state of a pinwheel type was experimentally observed at low temperatures $1.7K$, and again no magnetic order was seen over a wide range of higher temperatures ($J\sim 180$K with $D/J \approx 0.2$).\cite{pinwheel}  Finally, the material Cs$_{2}$Cu$_{3}$CeF$_{12}$ is a buckled kagome lattice with a quasi-one-dimensional structure in the exchange constants (similar to that shown in Fig.~\ref{fig:kagome}) which shows clear signs of order at low temperature.\cite{exp1} Taken together, these data indicate that while these systems may appear to be heading towards quantum spin liquid states over a wide temperature range, weak interactions, such as spin-orbit coupling, can eventually select an ordered phase.\cite{Starykh:prb10}  It is important to note that in real materials (which are three dimensional) the DM interaction may take a slightly different form than the strictly two-dimensional form we study here.  However, provided the layers are sufficiently weakly coupled on the energy scales for which we find ordering, our results should remain unchanged.

\subsection{Overview of theoretical approach and summary of results}

In this paper we focus on the spatially anisotropic spin-1/2 kagome lattice model.  Previous studies  in this limit have neglected the DM interaction.\cite{Oleg4,Apel1,Apel2,Wang:prb07,Nakano}
In our work, we closely follow the  quasi one-dimensional approach developed in Refs.[\onlinecite{Oleg1,Oleg2,Oleg3,Oleg4}] based on the application of the powerful one-dimensional (non-Abelian) bosonization technique.\cite{Giamarchi, Gogolin} In this approach one constructs a two-dimensional system by weakly coupling one-dimensional spin chains together.  Thus, the exchange interaction between spin chains is assumed smaller than the exchange interaction between spins on a given spin chain. (See Fig.\ref{fig:kagome}.)  In a one-dimensional spin-$1/2$ chain, the spin degrees of freedom can be represented in the spin-current formalism (non-Abelian bosonization) and the low-energy theory of the spin chains is described by a $SU(2)_{1}$ Wess-Zumino-Novikov-Witten (WZNW) theory.\cite{Gogolin} The inter-chain interactions can then be analyzed by perturbation theory using a scaling and a renormalization group approach.  The basic structure that emerges from this analysis is the following:\cite{Oleg4} there are two temperature scales that enter into the magnetic ordering. First, the middle spins order in either a  spiral state with small wave vector or a ferromagnetic state (spiral state with vanishing ordering vector) on a temperature scale $T_m\sim (J')^2/J$. As the temperature decreases further the spin chains also begin to order on a scale $T_{ch}\sim (J')^4/J^3$, with an order dictated by the magnetic state of the middle spins.\cite{Oleg4} Because of this separation of energy scales, the middle spins produce an effective exchange field on the spins of the spin chains which ultimately plays an important role in the magnetic order at the lowest temperatures.  We note that we do not find a spin-liquid ground state in any regime we study.

Our main results are the following. For weak spin-orbit coupling, the DM interaction selects the plane of the kagome lattice in which the middle spins take on a spiral order (the plane of the kagome lattice itself).  Our analysis of the spin chains shows that the spins along a chain are predominantly anti-parallel to the direction of the middle spins. However, there is a small component which is perpendicular to the main order, and this component undergoes a phase transition as the DM interaction is increased.   For small values of the DM interaction the perpendicular component orders antiferromagnetically perpendicular to the plane of the spiral, while for larger values this component becomes a mixture of coexisting antiferromagnetic and dimerized states and rotates into the plane of the spiral.  We estimate the value of the critical DM interaction [see Eq.\eqref{crit_cond}] to be $D/J \sim (J'/J)^{5/2}$.

The remainder of the paper is organized as follows. In Sec~\ref{sec:model} we begin with a tight-binding model with spin-orbit coupling on the kagome lattice. We then derive the effective spin Hamiltonian in the limit of a large on-site repulsion at half-filling. Thus, the parameters of a spatially anisotropic kagome antiferromagnet are determined.  In this section, a non-Abelian bosonization mapping is introduced that we will use throughout the remainder of the paper. In Sec.~\ref{sec:perturbation} we perform perturbation theory in the exchange interaction $J'$ between spin chains and middle spins.  Our approach closely follows that of Ref.[\onlinecite{Oleg4}].   In Sec.~\ref{sec:order} the order of the spin chains and the two dimensional system is analyzed from the point-of-view of perturbative renormalization group equations.  Finally, we present our main conclusions in Sec.~\ref{sec:conclusions}.  Some technical formulas are given in Appendix~\ref{app:fusion} and Appendix~\ref{app:RG_second}.

\section{Model Hamiltonian}
\label{sec:model}
In order to remind the reader of the microscopic origin of the DM interaction, and to establish some relationship with other phases where spin-orbit coupling plays a role (such as topological band insulators\cite{Moore:nat10,Hasan:rmp10,Qi:rmp11}), we first consider a tight-binding model on the {\em spatially isotropic} kagome lattice with intrinsic spin-orbit coupling.  We study a model with only on-site interactions, and assuming half-filling  we derive an effective spin Hamiltonian. The spin-orbit coupling translates into a DM interaction which induces an anisotropy in the exchange interaction. With insight from the spatially isotropic case, we then describe the case of a spatially anisotropic kagome lattice, which is the main focus of this paper.

\subsection{Tight-binding model on the kagome lattice with spin-orbit coupling}
In this section we give a derivation of the nearest-neighbor spin Hamiltonian in the presence of spin orbit-coupling on the kagome lattice. The calculation is standard,\cite{auerbach, aharony} but clearly illustrates how the DM interaction is obtained and gives the precise form we will use in this work.

For a one-electron tight-binding model on the kagome lattice the nearest neighbor hopping Hamiltonian is given by
\begin{equation}\label{tight-binding}
H_{t}=-t\sum_{\langle ij\rangle \alpha}c^{\dag}_{i\alpha}c_{j\alpha},
\end{equation}
where $t$ is the hopping, and  $c_{i\alpha}$ is an operator that annihilates a fermion on a site $i$ with spin $\alpha$. Likewise, $c_{i\alpha}^{\dag}$ is the analogous fermion creation operator.  The notation $\langle..\rangle$ denotes nearest neighbor sites, and will be used throughout the paper.

The symmetry of the kagome lattice allows a nearest-neighbor spin-orbit hopping term of the form
\begin{equation}\label{tight-binding-so}
H_{so}=i\lambda_{so}\sum_{\langle ij\rangle \alpha\beta}\nu_{ij}c^{\dag}_{i\alpha}\sigma^{z}_{\alpha\beta}c_{j\beta},
\end{equation}
where $\lambda_{so}$ parameterized the spin-orbit coupling strength, and $\nu_{ij}=\pm$ depending on the direction of hopping--``+'' when the third site in a triangle is ``to the left" and ``-" when the third site in a triangle is ``to the right" (see Fig.~\ref{fig:kagome}).\cite{Kane:prl05,Kane_2:prl05}  Here $\sigma^z$ is the Pauli spin matrix representing the $z$-component of the spin.

We are interested in the strongly interacting limit at half-filling, which is a Mott insulator.  We model the interactions by a Hubbard on-site repulsive potential
\begin{equation}
H_{U}= U\sum_{i}n_{i\uparrow}n_{i\downarrow},
\end{equation}
where $n_{i\alpha}=c^{\dag}_{i\alpha}c_{i\alpha}$. In the limit of $t \ll U$, we perform perturbation theory in $t$ and obtain the spin Hamiltonian with anisotropic spin exchange (even in the present limit of an isotropic lattice),
\begin{eqnarray}\label{hmagnetic}
&H_{M}=J\sum_{\langle ij\rangle} \left[ \cos(2\theta)(S_{i}^{x}S_{j}^{x} + S_{i}^{y}S_{j}^{y}) + S_{i}^{z}S_{j}^{z} \right. \nonumber\\ &\left.
-\nu_{ij}\sin(2\theta)[{\bf S}_{i}\times {\bf S}_{j}]_{z}  \right],
\end{eqnarray}
where
\begin{equation}
\label{eq:J}
J=\frac{4(t^{2}+\lambda_{so}^{2})}{U},
\end{equation}
 and
 \begin{equation}
 \label{eq:theta}
 \theta=\arctan(\lambda_{so}/t).
 \end{equation}
 The first line in Eq.\eqref{hmagnetic} is the weakly anisotropic (when $\lambda_{so} \ll t$) exchange coupling between nearest-neighbor spins, and the second line is a staggered DM interaction with strength,
 \begin{equation}
 \label{eq:DM}
 D^z_{ij}=-J\nu_{ij}\sin(2\theta).
 \end{equation}
 In this paper we are going to focus only on this DM interaction, which is an intrinsic property of the two-dimensional kagome lattice.
 We would like to remark that another way to obtain the results described above is to note that the spin-orbit coupling can be absorbed in to a definition of the first-nearest neighbor hopping matrix element $t$, making it spin-dependent.\cite{aharony} The equations (\ref{hmagnetic})-(\ref{eq:theta}) describe the spin Hamiltonian of the {\em spatially isotropic} kagome lattice. When the lattice is anisotropic, the form is slightly modified as we describe below.

\subsection{Spatially anisotropic case}
In the previous section we derived the Hamiltonian of spins on the isotropic kagome lattice, and found the DM conventions are of the same form earlier studied in the literature.\cite{Sachdev2,Cepas1,Cepas2} Let us now describe how the interactions are modified in the spatially anisotropic case which will be studied in the remainder of the paper.

In a spatially anisotropic lattice such as that shown in Fig.~\ref{fig:kagome}, the bonds between horizontal chains are ``stretched" leading to a modified hopping $t'<t$ (and spin-orbit coupling) along the diagonals.  (In fact, the spin-orbit coupling will also be modified along the chains as well because its value is dependent on the position of the ``middle spins".)  The modified hopping parameters and spin-orbit coupling will change the exchange constants in \eqref{eq:J} and \eqref{eq:DM} along the diagonal bonds.  We take the exchange between spins on diagonals to be $J^{\prime}$ and on horizontal lines to be $J$, as before.  The DM interaction also takes on a similar anisotropy with an angle $\theta$ described as in \eqref{eq:theta} (slightly modified from the isotropic case since the spin-orbit coupling is modified, but this will not play a crucial role in our study) along the chains, and a similarly defined $\theta^{\prime}$ (in terms of $t'$ and $\lambda_{so}'$) along diagonals, giving
 \begin{equation}
 \label{eq:DM_p}
 D^{z'}_{ij}=-J'\nu_{ij}\sin(2\theta').
 \end{equation}

Thus, in the spatially anisotropic case the Hamiltonian has the same form as \eqref{hmagnetic}, only with different exchange and DM interactions along chains and between spins on the chains and the middle spins.  Since we are interested in the limit $J' \ll J$, we will first focus on the Hamiltonian of the spin chains and later treat the interaction with middle spins perturbatively. The spin chain Hamiltonian is thus
\begin{eqnarray}\label{h0}
& H_{0}=J\sum_{\langle ij\rangle , y}\left[\cos(2\theta)(S_{i,y}^{x}S_{j,y}^{x} + S_{i,y}^{y}S_{j,y}^{y}) + S_{i,y}^{z}S_{j,y}^{z} \right] \nonumber\\
& +J\sin(2\theta)\sum_{\langle ij\rangle ,y}\nu^{y}_{ij}[{\bf S}_{i,y}\times {\bf S}_{j,y}]_{z},
\end{eqnarray}
with $J>0$ given by \eqref{eq:J} as before, the index $y$ labels the different spin chains, and $\nu^{y}_{ij} = \pm 1$ which alternates from bond to bond. For even and odd spin chains the $\nu^{y}_{ij}$ differs by a sign for the same $\langle ij \rangle$ bond. These features are evident from Fig.~\ref{fig:kagome}.

We now absorb the DM term into the first term of \eqref{h0} by rotating spins with a {\em local} spin rotation, \cite{Affleck}
\begin{equation}\label{rotation0}
{\hat R}_{DM}(\theta) = \left(\begin{array}{ccc}
\cos(\theta) & -\sin(\theta) & 0 \\
\sin(\theta) & \cos(\theta) & 0 \\
0 & 0 & 1
\end{array}\right),
\end{equation}
which rotates the spin on site $i$ on chain $y$ as ${\bf S}_{i,y} \rightarrow {\hat R}_{DM}\left((-1)^{i+y+1}\theta\right){\bf S}_{i,y}$.
After the rotation, the Hamiltonian (\ref{h0}) becomes an isotropic Heisenberg model
\begin{equation}\label{spinchains}
H_{0} = J\sum_{\langle ij \rangle ,y}{\bf S}_{i,y}\cdot{\bf S}_{j,y},
\end{equation}
which is very convenient for performing perturbation theory in the interactions with middle spins. For a purely one dimensional system, this result of absorbing a staggered DM interaction was obtained in Ref.[\onlinecite{aharony}], which also emphasized that the DM interaction can play an essential role in spin ordering when frustration is present. As  we now show, that general result also applies to the model we study.

We start with the interactions between the spin chains with middle spins (those lying along the diagonals in Fig.~\ref{fig:kagome} between the spin chains). These terms consist of XXZ-type terms  and the DM interaction. Performing the rotation \eqref{rotation0} on the spin chains as described above, we arrive at the following form of the interaction
\begin{widetext}
\begin{eqnarray}\label{interaction}
&V = J^{\prime}\cos(\chi)\sum_{i,y} s^{x(y)}_{2i \pm 1/2,2y \mp 1/2}\left[ S^{x(y)}_{2i,2y \mp 1} + S^{x(y)}_{2i \pm 1,2y \mp 1} + S^{x(y)}_{2i,2y} + S^{x(y)}_{2i \pm 1,2y} \right] \nonumber\\
&+ J^{\prime} \sum_{i,y} s^{z}_{2i \pm 1/2,2y \mp 1/2}\left[ S^{z}_{2i,2y \mp 1} + S^{z}_{2i \pm 1,2y \mp 1} + S^{z}_{2i,2y} + S^{z}_{2i \pm 1,2y} \right] \nonumber\\
&+J^{\prime}\sin(\chi)\sum_{i,y}\left[ - [{\bf s}_{2i \pm 1/2,2y \mp 1/2} \times {\bf S}_{2i,2y\mp 1}]_{z}  + [{\bf s}_{2i \pm 1/2,2y \mp 1/2} \times {\bf S}_{2i \pm 1,2y\mp 1}]_{z} \right. \nonumber\\ &\left.
+[{\bf s}_{2i \pm 1/2,2y \mp 1/2} \times {\bf S}_{2i,2y}]_{z} - [{\bf s}_{2i \pm 1/2,2y \mp 1/2} \times {\bf S}_{2i \pm 1,2y}]_{z} \right],
\end{eqnarray}
where $\chi = 2\theta^{\prime} + \theta$ and the small ``s" describes the spin degrees of freedom of the middle spins.  Note that the first two terms describe the nearly isotropic exchange interaction (for weak spin-orbit coupling), while the last two terms describe the DM interaction between middle spins and the spins on the spin chains.

Focusing on the case, $J^{\prime} \ll J$, we can apply powerful bosonization methods to describe the low-energy physics of spin chains and construct perturbation theory in the interaction inter-chain interactions.  The basic approach has been successfully applied before to anisotropic systems {\em without spin-orbit coupling}.\cite{Oleg4} We begin by taking the continuum limit of local spin degrees of freedom which leads to a sum of a uniform and staggered magnetization,\cite{Oleg4}
\begin{equation}
\label{eq:low_energy}
{\bf S}_{i} \rightarrow a_{0}\left[ {\bf M}(x) + (-1)^{x}{\bf N}(x) \right],
\end{equation}
where $x=ia_{0}$, with $a_{0}$ being the lattice spacing, and $i$ labels sites along the chains. The definitions of these operators and the fusion rules which they obey are heavily discussed in the literature.\cite{Oleg1,Oleg2,Oleg3,Oleg4} For completeness,  we provide the necessary details in Appendix \ref{app:fusion}. In this representation the Hamiltonian \eqref{spinchains} is described by the WZNW SU$(2)_{1}$ theory.\cite{Gogolin}  Substituting the low-energy form \eqref{eq:low_energy} into (\ref{interaction}) we find,
\begin{eqnarray}\label{int}
& V_{1} = {\tilde \gamma}_{1} \sum_{x,y}  s^{x(y)}_{2x \pm 1/2, 2y \mp 1/2} \left( M^{x(y)}_{2y \mp 1}(2x) + M^{x(y)}_{2y}(2x)\right)
         +\gamma_{1} \sum_{x,y} s^{z}_{2x \pm 1/2, 2y \mp 1/2} \left( M^{z}_{2y \mp 1}(2x) + M^{z}_{2y}(2x)\right), \nonumber \\
& V_{2} = \gamma_{2}\sum_{x,y} \left[ -s^{x}_{2x \pm 1/2,2y \mp 1/2} ( N^{y}_{2y}(2x) - N^{y}_{2y \mp 1}(2x) ) +
                                                             s^{y}_{2x \pm 1/2,2y \mp 1/2} ( N^{x}_{2y}(2x) - N^{x}_{2y \mp 1}(2x) )\right], \\
& V_{3} = {\tilde \gamma}_{3}\sum_{x,y} (\pm)s^{x(y)}_{2x \pm 1/2,2y \mp 1/2} \partial_{x}( N^{x(y)}_{2y}(2x) + N^{x(y)}_{2y \mp 1}(2x) )
          +\gamma_{3}\sum_{x,y} (\pm)s^{z}_{2x \pm 1/2,2y \mp 1/2} \partial_{x}( N^{z}_{2y}(2x) + N^{z}_{2y \mp 1}(2x) ), \nonumber
\end{eqnarray}
\end{widetext}
where the last interaction, $V_3$, is important as it will generate relevant inter-chain couplings in the renormalization group analysis to follow.\cite{Oleg4}

In \eqref{int} we have used the approximations $M(2x+a_{0}) \approx M(2x)$ and $N(2x+a_{0}) \approx N(2x) + \frac{a_{0}}{2} \partial_{x} N(2x)$. The coupling constants in \eqref{int},  $\gamma_{1}=2a_{0}J^{\prime}$, ${\tilde \gamma}_{1}=2a_{0}J^{\prime}\cos(\chi)$, $\gamma_{2}=2a_{0}J^{\prime}\sin(\chi)$, $\gamma_{3}=-\frac{a_{0}^{2}}{2}J^{\prime}$, and ${\tilde \gamma}_{3}=-\frac{a_{0}^{2}}{2}J^{\prime}\cos(\chi)$ will flow under the renormalization group. Since $\chi= 2\theta'+\theta$, the strength of the spin-orbit coupling terms partially set the scale of the initial coupling constants.  We now proceed to treat the interaction  $V=V_{1}+V_{2}+V_{3}$ in perturbation theory.

Note that $V_2 \equiv 0$ if the DM interaction vanishes.  The term $V_2$ will turn out to give rise to dimer correlations for sufficiently large DM interaction, and therefore may potentially drive a quantum phase transition.

\section{Perturbation theory and the ordering of the middle spins}
\label{sec:perturbation}
The remainder of the paper is devoted to an analysis of the phase diagram of the anisotropic kagome antiferromagnet with spin-orbit coupling based on the perturbative interactions in \eqref{int}.  The basic elements of our approach closely follow that of  Ref.[\onlinecite{Oleg4}].  The new element in the present study is the presence of a DM interaction.

The low-energy form of the spin chain Hamiltonian (\ref{spinchains}) can be written using non-Abelian bosonization (spin current operators)\cite{Oleg4} as a Sugawara Hamiltonian with a backscattering term\cite{Gogolin}
\begin{equation}\label{sugawara}
H_{0}=\frac{2\pi u}{3}\int dx \left[ {\bf J}_{R}\cdot{\bf J}_{R} + {\bf J}_{L}\cdot{\bf J}_{L} \right] + g_{bs}\int dx~ {\bf J}_{R}\cdot{\bf J}_{L},
\end{equation}
where $u=\pi Ja_{0}/2$, and the second term in this expression is the backscattering with strength $g_{bs}<0$.\cite{Oleg4} In this section the backscattering will not play any role. However, it will be important in subsequent sections.

The operators ${\bf M}$ and ${\bf N}$ [from Eq.\eqref{eq:low_energy}] that enter expression (\ref{int}) have scaling dimension $1$ and $1/2$, respectively. That is, under rescaling of coordinate $x_{\ell}=xe^{-l}$ and time $\tau_{\ell}=\tau e^{-l}$, where $b=e^{d\ell}$ they change as
\begin{eqnarray}
M_y(x,\tau) &=& b^{-1}M_y(x/b,\tau/b), \\
N_y(x,\tau) &=& b^{-1/2} N_y(x/b,\tau/b).
 \end{eqnarray}
There is another important operator $\epsilon_y(x, \tau)$, the dimerization operator, related to the continuum limit of the scalar product of two neighboring spins, ${\bf S}_{i,y}\cdot {\bf S}_{i+1,y} \to (-1)^x\epsilon_y(x)$, which does not appear in the interactions (\ref{int}) but will be generated under the renormalization group flow.\cite{Oleg4}  It has scaling dimension $1/2$, so changes as $\epsilon_y(x,\tau) = b^{-1/2} \epsilon_y(x/b,\tau/b)$ under a rescaling of space and time. All of these operators obey fusion rules (see Appendix \ref{app:fusion}) which will be used in perturbation theory beyond leading order.

A symmetry analysis yields the form of all the possible terms that are expected to be generated through the renormalization group (RG) transformations.\cite{Oleg4} The most relevant ones, with lowest scaling dimension, are\cite{Oleg4}
\begin{eqnarray}
\label{interchainrelevant}
H_{N}=\sum_{y}\left[\gamma_{N1}\int dx N^{z}_{y}N^{z}_{y+1} + \gamma_{N2}\int dx N^{x(y)}_{y}N^{x(y)}_{y+1} \right] ,\nonumber\\
 H_{\epsilon} =\sum_{y} \gamma_{\epsilon}\int dx \epsilon_{y}\epsilon_{y+1}. \hspace{2cm}
\end{eqnarray}
There are also irrelevant interactions between spin chains which are important because they will {\em generate relevant terms} of the form \eqref{interchainrelevant}
\begin{eqnarray}
\label{interchainrelevant_gen}
H_{\partial N}& =& \sum_y\gamma_{\partial N1}\int dx \partial_{x}N^{z}_{y}\partial_{x}N^{z}_{y+1}\nonumber \\
&&+\sum_y \gamma_{\partial N2}\int dx \partial_{x}N^{x(y)}_{y}\partial_{x}N^{x(y)}_{y+1}, \nonumber\\
 H_{M}&=&\sum_y\gamma_{M1}\int dx M^{z}_{y}M^{z}_{y+1}\nonumber \\
 &&+ \sum_y\gamma_{M2}\int dx M^{x(y)}_{y}M^{x(y)}_{y+1}.
\end{eqnarray}

With \eqref{interchainrelevant} and \eqref{interchainrelevant_gen} in hand as ``targets" for important terms to be generated by the RG, we proceed to apply perturbation theory to second order in interaction between the spin chains and middle spins,
\begin{equation}
\label{eq:partition}
Z=\int e^{-S_{0}}\left[1-\int d\tau V(\tau) + \frac{1}{2}T\int d\tau_{1}d\tau_{2}V(\tau_{1})V(\tau_{2})  \right],
\end{equation}
where $V=V_{1}+V_{2}+V_{3}$ is given by \eqref{int}. In \eqref{eq:partition} $S_{0}$ is the fixed-point action of spin chains, and $T$ is the time ordering operator.

At second order in the interaction, the following combinations of middle spins will occur:
$T s^{a}_{i,y}s^{b}_{j,y^{\prime}}$. For the same site ($i=j$, $y=y^{\prime}$) the relationship $T s^{a}(\tau_{1})s^{b}(\tau_{2}) = \frac{1}{4}\delta^{ab}+\frac{i}{2}[\theta (\tau_{1}-\tau_{2}) - \theta(\tau_{2}-\tau_{1}) ]\epsilon^{abc}s^{c}((\tau_{1}+\tau_{2})/2)$ is obeyed, where $\epsilon^{abc}$ is the fully antisymmetric tensor, and $\theta(\tau)$ is the heaviside function: $\theta(\tau)=1$ for $\tau>0$ and  $\theta(\tau)=0$ for $\tau<0$. To derive the inter-chain operators one has to pick middle spins on the same sites with the same components (so that the middle spin degree of freedom ``drops out"). Integration over the relative time is performed over the interval $\alpha<|\tau_{1}-\tau_{2}|<b\alpha$ with $\alpha=a_0/u$.

To obtain contributions to the interaction $V$ in \eqref{int} one must select middle spins on the same site but with different components.  After straightforward but somewhat tedious calculations, we obtain the RG equations
\begin{eqnarray}
\label{eq:RG}
&\frac{d{\tilde \gamma}_{1}}{d\ell}=\frac{{\tilde \gamma}_{1}\gamma_{1}}{2\pi u}, ~~\frac{d\gamma_{1}}{d\ell}=\frac{{\tilde \gamma}_{1}^{2}}{2\pi u}, ~~\frac{d\gamma_{2}}{d\ell}=\frac{1}{2}\gamma_{2}+\frac{\gamma_{1}\gamma_{2}}{4\pi u},  \nonumber\\
&\frac{d{\tilde \gamma}_{3}}{d\ell}=-\frac{1}{2}{\tilde \gamma}_{3}+\frac{\gamma_{1}{\tilde \gamma}_{3}}{\pi u},~~\frac{d\gamma_{3}}{d\ell}=-\frac{1}{2}\gamma_{3}+\frac{{\tilde \gamma}_{1}{\tilde \gamma}_{3}}{\pi u}, \nonumber\\
&\frac{d\gamma_{\epsilon}}{d\ell} = \gamma_{\epsilon} - \frac{(\gamma_{\partial N1}\gamma_{M1}+2\gamma_{\partial N2}\gamma_{M2} )}{8\pi u^{3}\alpha^{2}}, \\
&\frac{d\gamma_{N2}}{d\ell} = \gamma_{N2} + \frac{\gamma_{2}^{2}}{4u} + \frac{\gamma_{M1}\gamma_{\partial N2}}{8\pi u^{3}\alpha^{2}}, \nonumber\\
&\frac{d\gamma_{N1}}{d\ell} = \gamma_{N1} + \frac{\gamma_{M2}\gamma_{\partial N1}}{8\pi u^{3}\alpha^{2}}, \nonumber\\
&\frac{d\gamma_{M2}}{d\ell} = \frac{\gamma_{M1}\gamma_{M2}}{4\pi u} - \frac{{\tilde \gamma}_{1}^{2}}{4u}, ~~\frac{d\gamma_{M1}}{d\ell} = \frac{\gamma_{M2}^{2}}{4\pi u} - \frac{\gamma_{1}^{2}}{4u}, \nonumber\\
&\frac{d\gamma_{\partial N2}}{d\ell} = -\gamma_{\partial N2} - \frac{{\tilde \gamma}_{3}^{2}}{4u}, ~~~\frac{d\gamma_{\partial N1}}{d\ell} = -\gamma_{\partial N1} - \frac{\gamma_{3}^{2}}{4u},\nonumber
\end{eqnarray}
which are the generalization of the corresponding results in Eq.(14) of Ref.[\onlinecite{Oleg4}].  (We find a small difference in the numerical factors of the second order terms compared to Ref.[\onlinecite{Oleg4}], but they do not affect any of our overall conclusions or those of that work. An example derivation is given in Appendix B.)

We also obtained RG equations for exchange coupling constants of interaction between middle spins. We are not showing them here since they resemble those given in Ref.[\onlinecite{Oleg4}]. The only modification due to DM interaction is the parametric enhancement of exchange in $x - y$ plane.

While the equations \eqref{eq:RG} are somewhat more involved than  the case of vanishing DM interaction, they may be analyzed in a similar fashion.  Because the DM interaction is expected to be small compared to $J',J$ (or at most of the same order), the hierarchy of energy scales at higher energies is not expected to change.  Namely, there an ordering temperature for the middle spins $T_{m} \propto (J^{\prime})^{2}/J$, which should be relatively insensitive to the DM interactions.  The spin chains will start to order only at a lower temperature $T_{ch} \propto (J^{\prime})^{4}/J^{3}$.\cite{Oleg4} The procedure for determining the low-temperature magnetic order (that is, below $T_{ch}$) is to first find the order of the middle spins.  These middle spins will then produce an average exchange field that acts on the spin chains.  Thus, the final ordering of the anisotropic kagome lattice will be obtained by analyzing the spin chains in the presence of a (potentially) spatially varying exchange field.

At the energy scale for which the middle spins interactions become non-perturbative, corresponding to $\ell \sim 1$,
one can estimate the values of the relevant coupling constants of inter-chain interactions,
\begin{eqnarray}
& \gamma_{\epsilon} \sim -\frac{a_{0}(J^{\prime})^{4}}{16 \pi^{4} J^{3}}(1+2\cos^{4}(\chi)), ~~ \gamma_{N1} \sim \frac{a_{0}(J^{\prime})^{4}}{16 \pi^{4} J^{3}}\cos^{2}(\chi) \nonumber\\
& \gamma_{N2} \sim \frac{a_{0}(J^{\prime})^{4}}{16 \pi^{4} J^{3}}\cos^{2}(\chi) + \frac{2a_{0}(J^{\prime})^{2}}{\pi J}\sin^{2}(\chi),
\end{eqnarray}
where we have used that $\gamma_{\partial N 1}\sim - \frac{\gamma_{3}^{2}}{4u}$, $\gamma_{\partial N 1}\sim - \frac{{\tilde \gamma}_{3}^{2}}{4u}$, $\gamma_{M1} \sim - \frac{{\tilde \gamma}_{1}^{2}}{4u}$,  $\gamma_{M2}\sim  - \frac{\gamma_{1}^{2}}{4u}$, and substituted in the corresponding ``initial" values of $\gamma_1,\tilde\gamma_1,\gamma_2,\gamma_3,\tilde \gamma_3$ given below Eq.\eqref{int}.

In order to determine the order of the middle spins (which form a triangular lattice) we follow arguments of Ref.[\onlinecite{Oleg4}] which uses perturbation theory to derive the effective Hamiltonian of interactions between middle spins.  Before we proceed, however, one comment is in order.  A full solution of the effect of the DM interactions on the middle spins is involved.  As we have argued earlier, the DM interaction is not expected to have a significant effect on the ordering temperature, $T_{m}$.  However, it still plays a role:  It selects the plane of the spiral order found in  Ref.[\onlinecite{Oleg4}] to be the $x-y$ plane.  (Without the DM interaction, the plane of the spiral order is arbitrary.)  With the plane of the spiral order determined, we then follow Ref.[\onlinecite{Oleg4}] and write the Hamiltonian of the middle spins as
\begin{eqnarray}\label{H_triangle}
&H_{\triangle} = 2(J^{\prime})^{2} \left( A(1) - B(1) \right) \sum_{\langle ij \rangle} s^{x(y)}_{i}s^{x(y)}_{j}  \nonumber \\
& + 4(J^{\prime})^{2} \left( A(2) - B(2) \right) \sum_{[ij]} s^{x(y)}_{i}s^{x(y)}_{j},
\end{eqnarray}
where notation $\langle..\rangle$ denotes first nearest neighbor, and $[..]$ second nearest neighbor pairs of spins. The coefficients of these terms involve
\begin{eqnarray}
&A(r) = \cos^{2}(\chi)\left(2G_{M}(r) + G_{M}(r+1) + G_{M}(r-1) \right), \nonumber \\
&B(r) = \sin^{2}(\chi)\left(2G_{M}(r) - G_{M}(r+1) - G_{M}(r-1) \right), \nonumber
\end{eqnarray}
where
\begin{equation}
\label{G}
G_{M}(r) = \frac{2}{\pi} \int_{0}^{\infty} d\omega^{\prime}\int_{0}^{\pi}dq~ S(q,\omega^{\prime}) \frac{\cos(qr)}{\omega^{\prime}},
\end{equation}
and $S(q,\omega^{\prime})$ is the dynamical structure factor. With the definition \eqref{G}, we find
\begin{eqnarray}
\label{eq:A-B}
&A(r)-B(r) = \frac{8}{\pi} \int_{0}^{\infty} d\omega^{\prime}\int_{0}^{\pi}dq~ \cos^{2}(q/2)\cos(qr)  \frac{S(q,\omega^{\prime})}{\omega^{\prime}}\nonumber\\
& - \frac{8\sin^{2}(\chi)}{\pi} \int_{0}^{\infty} d\omega^{\prime}\int_{0}^{\pi}dq~ \cos(qr)  \frac{S(q,\omega^{\prime})}{\omega^{\prime}}.
\end{eqnarray}
The first integral in \eqref{eq:A-B} was numerically estimated in the work Ref.[\onlinecite{Oleg4}], and it scales as $1/J$.  At zero temperature, the second integral in \eqref{eq:A-B} diverges, but
for finite temperatures small compared to $J$, this integral scales as $1/T$.\cite{structure1, structure2}   Thus, overall one has that the exchange coupling in \eqref{H_triangle}, $J_{eff}= (J'^2)\left[  \frac{c_1}{J}+\frac{c_2\sin^{2}(\chi)}{T}\right]$, where $c_1,c_2$ are order one constants.  Because the middle spins order at a finite temperature given approximately by $T_m \sim (J^{\prime})^{2}/J$, the ratio of the second integral to the first integral in \eqref{eq:A-B} can be estimated as $J \sin^{2}(\chi)/T_m \sim (J\sin(\chi)/J^{\prime})^{2}$, which we assume to be smaller than one.

The result of this analysis is that for temperatures of order $T_{crit} \sim J \sin^{2}(\chi)$ and larger, we may use the results of Ref.~[\onlinecite{Oleg4}] for the ordering of middle spins as a good approximation. Classical considerations and an analysis of the two-spinon dynamical structure factor suggest that spiral order is a leading candidate,\cite{Oleg4}
\begin{equation}\label{spiral}
\langle{\bf s}_{x}\rangle = s_{0}[{\bf e}_{x}\cos(qx) + {\bf e}_{y}\sin(qx)],
\end{equation}
where ${\bf e}_{x(y)}$ is a unit vector along the $x(y)$ coordinate, and with $q\ll1$, and $q=0$ (ferromagnetic) a distinct possibility. Here $s_0\lesssim 1/2$ is the local static moment of the middle spins.  Below we study the effect of these two possible middle spin orders on the ordering of the spin chains: (i) spiral order in $x-y$ plane, and (ii) ferromagnetic order along $x$ direction.

We also showed that for small temperatures, when the second term in $J_{eff}$ dominates, the results of Ref. ~[\onlinecite{Oleg4}] still hold. Namely, the two-spinon dynamical structure factor approximation still predicts the spiral order of middle spins.

\section{Ordering of the spin chains in response to the ordering of the middle spins}
\label{sec:order}

We discuss how the spin chains order for $q$ finite and small, and $q=0$ in the middle spin ordering described in Eq.\eqref{spiral}.  At small Dzyaloshinskii-Moriya interaction strength we expect a spiral order (\ref{spiral}) with $q\ll 1$, since the results of Ref.~[\onlinecite{Oleg4}] remain unchanged above $T_{crit}$, and our estimates suggest that they hold below the $T_{crit}$. The ordered middle spins then produce an effective exchange field (which can be treated in mean-field theory) that acts on the spin chains and causes them to order.  We now determine the order of the spin chains on various temperature scales as a function of the Dzyaloshinskii-Moriya interaction.

To do so, we plug \eqref{spiral} back into (\ref{interaction}) and obtain the result that the order of the middle spins appears as an effective magnetic field for the spins on the chain,
\begin{eqnarray}\label{effectivemf}
&V^{odd/even}=J^{\prime}h\cos(\chi)\sum_{i}\left[\cos(qx)S^{x}_{i} + \sin(qx) S^{y}_{i} \right] \nonumber\\
&\pm J^{\prime}h\sin(\chi)\sum_{i}(-1)^{i}\left[\cos(qx)S^{y}_{i} - \sin(qx) S^{x}_{i} \right],
\end{eqnarray}
where $h=2s_{0}\cos(q/2)$ and the $x$ in the argument of the $\sin$ and $\cos$ is $x=i a_0$, with $a_0$ the spacing of the spins along the chains. The first line of \eqref{effectivemf} describes a uniform spiral magnetic field while the second line describes a staggered, slowly rotating (in space) magnetic field. Performing a local rotation of spins as ${\bf S}_{i} \to \hat R_{x}(x){\bf S}_{i}$, where
\begin{equation}\label{rotation1}
{\hat R}_{x}(x) = \left(\begin{array}{ccc}
\cos(qx) & \sin(qx) & 0 \\
-\sin(qx) & \cos(qx) & 0 \\
0 & 0 & 1
\end{array}\right),
\end{equation}
the Hamiltonian $H^{odd}=H_0+V^{odd}$ of the odd spin chains becomes,
\begin{eqnarray}\label{H_odd_rot}
&H^{odd}=J\cos(q)\sum_{i}{\bf S}_{i}\cdot{\bf S}_{i+1} +J (1-\cos(q))\sum_{i}S^{z}_{i}S^{z}_{i+1} \nonumber\\
&+J\sin(q)\sum_{i}\left(S^{y}_{i}S^{x}_{i+1}-S^{x}_{i}S^{y}_{i+1}\right) \nonumber\\
&+J^{\prime}h\sum_{i}\left( \cos(\chi)S_{i}^{x} + (-1)^{i}\sin(\chi) S_{i}^{y} \right),
\end{eqnarray}
where the first line describes anisotropic exchange between spins, the second line describes an effective {\em uniform} Dzyaloshinskii-Moriya interaction, and the third line are uniform and staggered magnetic fields resulting from exchange interactions with the middle spins.  For even chains, one may take $\chi \to -\chi$.  {\em Note that since the effect of the spin-orbit coupling is included in $\chi=2\theta+\theta'$, the main role of the DM interaction is to give rise to the staggered magnetic field in the last line of \eqref{H_odd_rot}.}

For the convenience of the following analysis we make a $\pi/2$ rotation about the $y$-axis which transforms $S^{x}_{i} \rightarrow S^{z}_{i}$ and $S^{z}_{i} \rightarrow -S^{x}_{i}$. We further assume that $q\ll1$ which allows us to neglect anisotropy $J(1-\cos(q))$ up to corrections of order $q^2$.

We now bosonize the spin chain spins in the non-Abelian spin-current representation. The Hamiltonian (\ref{sugawara}) of spin chains under the rotation (\ref{rotation1}) has the same form but with updated constants: $u \rightarrow \cos(q)u$, and $g_{bs} \rightarrow \cos(q)g_{bs}$. We note that the backscattering coupling constant is negative, $g_{bs} < 0$. The effective (uniform) Dzyaloshinskii-Moriya interaction (after the rotation about the $y$-axis) in \eqref{H_odd_rot} takes the form\cite{Suhas}
\begin{equation}\label{dm}
H^{\prime}=J\sin(q)\sum_{i}\left(S^{y}_{i}S^{z}_{i+1}-S^{z}_{i}S^{y}_{i+1}\right) = {\tilde d} \int~ dx (J_{R}^{x}-J_{L}^{x}),
\end{equation}
where ${\tilde d}=J\sin(q)\frac{4}{\pi}$. The uniform magnetic field (after the rotation about the $y$-axis) in \eqref{H_odd_rot} becomes
\begin{equation}\label{mf}
H^{\prime\prime}=J^{\prime}h\cos(\chi)\sum_{i}S_{i}^{z} = {\tilde h}\int~ dx (J_{R}^{z}+J_{L}^{z}),
\end{equation}
where ${\tilde h}=J^{\prime}h\cos(\chi)$, and the staggered magnetic field takes the form
\begin{equation}\label{st}
H_{st}=J^{\prime}h\sin(\chi)\sum_{i} (-1)^{i} S_{i}^{y} = \gamma_{n}\int~dxN^{y},
\end{equation}
where $\gamma_{n}=J^{\prime}h\sin(\chi)$. The corresponding Hamiltonian for the even spin chains is obtained by simply putting $\chi \rightarrow -\chi$ in expressions \eqref{mf} and \eqref{st} above.

We now perform a chiral rotation of the spin currents to absorb the Dzyaloshinskii-Moriya term (\ref{dm}) into the uniform magnetic field term, (\ref{mf}). The right and left spin currents can be rotated independently while keeping the WZNW field theory invariant. We make the rotation ${\bf J}_{R}\rightarrow {\hat R}_{R}{\bf J}_{R}$ and ${\bf J}_{L}\rightarrow {\hat R}_{L}{\bf J}_{L}$ using
\begin{equation}\label{rotation2}
{\hat R}_{R/L} = \left(\begin{array}{ccc}
\cos(\phi) & 0 & \mp\sin(\phi) \\
0 & 1 & 0 \\
\pm\sin(\phi) & 0 & \cos(\phi)
\end{array}\right),
\end{equation}
where $\phi = \arctan({\tilde d}/{\tilde h})$. Under this transformation the uniform Dzyaloshinksii-Moriya interaction and uniform magnetic field combine to take the form
\begin{equation}\label{mf_transformed}
H^{\prime}+H^{\prime\prime}=\sqrt{{\tilde h}^{2}+{\tilde d}^{2}}\int dx (J_{R}^{z} + J_{L}^{z}),
\end{equation}
and for convenience we define $h^{z} = \sqrt{{\tilde h}^{2}+{\tilde d}^{2}}$. The backscattering term transforms as
\begin{eqnarray}\label{backscattering_final}
&H_{bs}= g_{bs}\int dx\left(\frac{1}{4}(\cos(2\phi)-1)(J_{R}^{+}J_{L}^{+} + J_{R}^{-}J_{L}^{-} )  \right. \nonumber\\ &\left.
+ \frac{1}{4}(\cos(2\phi) + 1)(J_{R}^{-}J_{L}^{+} + J_{R}^{+}J_{L}^{-})  \right. \nonumber\\ &\left.
+ \frac{1}{2} (J_{R}^{z}J_{L}^{+} + J_{R}^{z}J_{L}^{-} - J_{R}^{+}J_{L}^{z} - J_{R}^{-}J_{L}^{z})\right. \nonumber\\ &\left.
+ \cos(2\phi)J_{R}^{z}J_{L}^{z} \right),
\end{eqnarray}
where $J^{\pm}=J^{x}\pm iJ^{y}$, and we define
\begin{eqnarray}
 g_{1}&=&\frac{1}{2}(\cos(2\phi)-1)g_{bs},\\
  g_{2}&=&\cos(2\phi)g_{bs},
  \end{eqnarray}
whose flow under the renormalization group will turn out to be important in the eventual analysis of the low-temperature phase of the spin chains.   The fields with scaling dimension $1/2$ transform as\cite{Oleg4}
\begin{eqnarray}
N^{x(z)} &\to & N^{x(z)},\\
 N^{y} &\to & \cos(\phi)N^{y} + \sin(\phi)\epsilon, \\
 \epsilon &\to &\cos(\phi) \epsilon - \sin(\phi) N^{y},
\end{eqnarray}
so that the staggered magnetic field becomes
\begin{equation}\label{staggerred_final}
H_{st} = \gamma_{n}\int dx \left( \cos(\phi)N^{y} + \sin(\phi) \epsilon \right).
\end{equation}
The inter-chain interactions transform as
\begin{eqnarray}\label{inter-rot}
&H_{N}+H_{\epsilon} = \int dx~\left( \gamma_{N1} N^{z}_{y}N^{z}_{y+1} + \gamma_{N2}N^{x}_{y}N^{x}_{y+1}   \right. \nonumber\\ &\left.
+ (\gamma_{N2}\cos^{2}(\phi) + \gamma_{\epsilon} \sin^{2}(\phi))N_{y}^{y}N_{y+1}^{y} \right. \nonumber\\ &\left.
+ (\gamma_{N2}\sin^{2}(\phi) + \gamma_{\epsilon} \cos^{2}(\phi))\epsilon_{y}\epsilon_{y+1}  \right. \nonumber\\ &\left.
+ \cos(\phi)\sin(\phi)(\gamma_{N2}-\gamma_{\epsilon})(N_{y}^{y}\epsilon_{y+1} + \epsilon_{y}N_{y+1}^{y})  \right).
\end{eqnarray}
The total Hamiltonian, which is our subject for further analysis, is a sum of (\ref{mf_transformed}), (\ref{backscattering_final}), (\ref{staggerred_final}), (\ref{inter-rot}), together with the free part expressed through spin-currents (\ref{sugawara}). In particular, the staggered magnetic field, Eq.\eqref{staggerred_final}, plays a key role in our result Eq.\eqref{result1}.
We now absorb the effective magnetic field (\ref{mf_transformed}) into the spin-currents by a shift of the bosonic field\cite{Oleg4} $\varphi_{s} \rightarrow \varphi_{s} +\frac{x h^{z}}{\sqrt{2\pi}u}$. The spin currents are then transformed as
\begin{eqnarray}\label{mfabsorb}
J_{R}^{\pm} = J_{R}^{\pm} e^{\mp ix h^{z}/u}, \nonumber\\
J_{L}^{\pm} = J_{L}^{\pm} e^{\pm ix h^{z}/u},\nonumber\\
J_{R/L}^z = J_{R/L}^z+\frac{h^{z}}{4 \pi u}.
\end{eqnarray}
The fields with scaling dimension $1/2$ transform as\cite{Oleg4}
\begin{eqnarray}\label{final-rot}
N^{z} = \cos\left(x h^{z}/u\right) N^{z} - \sin\left(x h^{z}/u\right) \epsilon, \nonumber\\
\epsilon = \cos\left(x h^{z}/u\right) \epsilon + \sin\left(x h^{z}/u\right) N^{z},\nonumber \\
\end{eqnarray}
and $N^{x,y}$ remain unchanged.  This shift renders the first and third terms in \eqref{backscattering_final} oscillatory on scales $x>u/h^{z}$ so that they will not contribute to the long distance, low-energy flows of the coupling constants.

\subsection{Analysis of the low-energy physics of the spin chains}

We will now use perturbation theory to study the coupling of the spin chains to the middle spins. We first note that there are two scales in the problem. At small length scales $\ell<\ell^{*}$, where $\ell^{*} \sim ln(J/J^{\prime})$ which is obtained as $a_{0}e^{\ell^{*}}h^{z}/u \sim 1$, oscillating factors do not play a role. One may assume that $h^{z}=0$ in the relations given immediately above. The coupling constants under renormalization in this case are given by
\begin{eqnarray}
& \frac{d g_{bs}}{d \ell} = \frac{g_{bs}^{2}}{2\pi u}, ~~~  \frac{d \gamma_{n}}{d \ell} = \left(\frac{3}{2} - \frac{g_{bs}}{4\pi u}  \right)\gamma_{n},\nonumber \\
& \frac{d \gamma_{N1(2)}}{d \ell} = \left(1 - \frac{g_{bs}}{4\pi u}  \right)\gamma_{N1(2)}, \nonumber\\
& \frac{d \gamma_{\epsilon}}{d\ell} = (1 + \frac{3g_{bs}}{4\pi u})\gamma_{\epsilon},
\end{eqnarray}
and should be stopped at $\ell\sim \ell^{*}$. Solving for the backscattering we get
\begin{equation}
g_{bs}(l)=\frac{g_{bs}^{0}}{1-\frac{\ell g_{bs}^{0}}{2\pi u}},
\end{equation}
which is a decreasing function of $\ell$ since  $g_{bs}^{0}<0$.\cite{Oleg4} The solutions for the coupling constants at $l \sim l^*$ are
\begin{eqnarray}\label{eq:short-flow}
&\gamma_{N1(2)} = \gamma_{N1(2)}^{0} e^{\ell^{*}}\left( 1 - \frac{\ell^{*} g_{bs}^{0}}{2\pi u} \right)^{1/2}, \nonumber\\
&\gamma_{\epsilon} = \gamma_{\epsilon}^{0} e^{\ell^{*}}\left( 1 - \frac{\ell^{*} g_{bs}^{0}}{2\pi u} \right)^{-3/2}, \nonumber\\
&\gamma_{n} = \gamma_{n}^{0} e^{3\ell^{*}/2} \left( 1 - \frac{\ell^{*} g_{bs}^{0}}{2\pi u} \right)^{1/2}.
\end{eqnarray}
The coupling constant $\gamma_{n}$ has the largest scaling dimension $3/2$ [from \eqref{st}]. We thus expect that it will grow at long length scales and compete with the inter-chain coupling constants, $\gamma_{N1(2)}$ (which have scaling dimension $1$), and result in a phase transition.

To study  the phase transition as a function of the Dzyaloshinskii-Moriya coupling we analyze renormalization group flows at large scales. At large scales $\ell > \ell^{*}$ rapid oscillations of various terms in the Hamiltonian [the first and third terms in \eqref{backscattering_final}] causes them to average to zero and we neglect them. The inter-chain interaction \eqref{inter-rot} after the rotation \eqref{final-rot} becomes
\begin{eqnarray}
&H_{N}+H_{\epsilon} = \int dx~ \left( \gamma_{x}N_{y}^{x}N_{y+1}^{x} + \gamma_{y}N_{y}^{y}N_{y+1}^{y} \right. \nonumber\\ &\left.
+ \gamma_{+}(\epsilon_{y}\epsilon_{y+1} + N_{y}^{z}N_{y+1}^{z}) \right),
\end{eqnarray}
where we have defined the coupling constants as $\gamma_{x}=\gamma_{N2}$, $\gamma_{y}=\gamma_{N2}\cos^{2}(\phi) + \gamma_{\epsilon} \sin^{2}(\phi)$, $\gamma_{+} = \frac{1}{2}(\gamma_{N2}\sin^{2}(\phi) + \gamma_{\epsilon} \cos^{2}(\phi) + \gamma_{N1})$. We only keep $g_{1}$ and $g_{2}$ terms in the backscattering (\ref{backscattering_final}), and only the first term, with $N^{y}$, in staggered magnetic field (\ref{staggerred_final}) and label its coupling constant ${\tilde \gamma}_{n} = \cos(\phi)\gamma_{n}$. The remaining terms are oscillatory and therefore ignored for this long-distance analysis.

The resulting renormalization group equations describing the coupling constants of the non-oscillatory terms are
\begin{eqnarray}
& \frac{d g_{1}}{d \ell} = - \frac{g_{1}g_{2}}{2\pi u}, ~~~ \frac{d g_{2}}{d \ell} = -\frac{g_{1}^{2}}{2\pi u}, \\
&\frac{d \gamma_{x}}{d \ell} = \left( 1 + \frac{g_{1}}{2\pi u} - \frac{g_{2}}{4\pi u} \right)\gamma_{x}, ~~~ \frac{d \gamma_{y}}{d \ell} = \left( 1 - \frac{g_{1}}{2\pi u} - \frac{g_{2}}{4\pi u} \right)\gamma_{y}, \nonumber\\
&\frac{d\gamma_{+}}{d\ell} = \left( 1 + \frac{g_{2}}{4\pi u} \right)\gamma_{+}, ~~~ \frac{d{\tilde \gamma}_{n}}{d\ell} = \left( \frac{3}{2} -\frac{g_{1}}{2\pi u} - \frac{g_{2}}{4\pi u}  \right){\tilde \gamma}_{n}.  \nonumber
\end{eqnarray}
The analysis\cite{Oleg4} of these equations show that there are two competing operators: $\gamma_{x}$ and ${\tilde \gamma}_{n}$. The latter is due to the Dzyaloshinkii-Moriya interaction (recall $\tilde \gamma_n=\cos(\phi)\gamma_n=\cos(\phi) J' 2 s_0 \cos(q/2)\sin( \chi)$ and $\gamma_x=\gamma_{N2}\sim(J')^4/J^3$ are the bare couplings before the short-scale renormalization).

From the short-scale renormalization, new ``bare" couplings emerge from which the long-scale renormalization flows begin.  These are determined from \eqref{eq:short-flow}.
 We give a rough estimate of the value of Dzyaloshinskii-Moriya interaction constant at which the phase transition between $\gamma_x$ and $\tilde \gamma_n$ dominated states occurs. Comparing lengths at which ${\tilde \gamma}_{n}e^{3\ell/2} \sim 1$ where ${\tilde \gamma}_{n} \sim \sin(\chi)\left(\frac{J}{J^{\prime}}\right)^{1/2}$ from  \eqref{eq:short-flow}, and $\gamma_{x}e^{\ell} \sim 1$ where $\gamma_{x} \sim \left(\frac{J^{\prime}}{J}\right)^{3}$ we get the critical value of the angle $\theta$ which defines the critical strength of Dzyaloshinkii-Moriya interaction in our model:
\begin{equation}\label{crit_cond}
\sin(\chi_{c})= \left(\frac{J^{\prime}}{J}\right)^{5/2},
\end{equation}
where for simplicity one may assume that $\theta \sim \theta^{\prime}$ since they are parametrically of similar order of magnitude,  giving $\chi_c \approx 3 \theta_c$.   The condition \eqref{crit_cond} is one of the central results of this work. For $\theta<\theta_{c}$ coupling constant $\gamma_{x}$ dominates, while for $\theta>\theta_{c}$ staggered magnetic field ${\tilde \gamma}_{n}$ dominates.

A rough numerical estimate of $\theta_c$ can be made since $\sin(2\theta)\approx D/J$ [from \eqref{eq:DM}], which gives $\sin(3\theta_c) \approx 3D/(2J) \approx (J'/J)^{5/2}$.  For $D/J \approx 0.2$ (possible in the class of materials described in the introduction), one finds the critical point would be reached for lattice sufficiently anisotropic that $J'/J \lesssim 0.7$.  While in experiment the precise value of $J'/J$ is not presently known for various anisotropic materials, it is possible that the application of pressure could further distort an anisotropic lattice to reach these values and possibly drive a phase transition. If $D$ is smaller, the critical ratio of $J'/J$ will in turn be smaller.  However, we expect our main conclusions to be qualitatively valid even for $J'/J \approx 0.7$, in spite of our initial assumption $J' \ll J$.

\subsection{Resulting order of the spin chains at low temperatures}
For $\theta>\theta_{c}$, one finds from \eqref{staggerred_final},  $\langle N^{y}_y \rangle = \cos(\phi) (-1)^{y}M_{1}$, where $M_{1} \sim \frac{J^{\prime}}{J}\sin(3\theta)$. Tracing back all the transformations performed (\ref{mfabsorb}), (\ref{rotation2}), (\ref{rotation1}), and (\ref{rotation0}), we obtain the final order of the spin chains,
\begin{eqnarray}\label{result1}
\langle S_{x,y}^{x} \rangle &=& -\frac{h^{z}}{2\pi u}\cos(\phi)\cos(qx - (-1)^{x+y}\theta) \nonumber\\
&& - (-1)^{x+y}M_{1}\cos(\phi)\sin(qx - (-1)^{x+y}\theta), \nonumber\\
\langle S_{x,y}^{y} \rangle &=& -\frac{h^{z}}{2\pi u}\cos(\phi)\sin(qx - (-1)^{x+y}\theta) \nonumber\\
&& + (-1)^{x+y}M_{1}\cos(\phi)\cos(qx - (-1)^{x+y}\theta),\nonumber\\
\langle S_{x,y}^{z} \rangle &=& 0,\nonumber \\
\langle \epsilon_y \rangle &=& -\sin(\phi) (-1)^{y}M_{1},
\end{eqnarray}
where $h^{z} = \sqrt{{\tilde h}^{2}+{\tilde d}^{2}}$ as given below \eqref{mf_transformed}.
This spin-orbit coupling dominated phase is characterized by a non-zero value of dimerization operator and staggered magnetization (Neel order) in the plane of the spiral (as opposed to Neel order in the plane perpendicular to the plane of the spiral as occurs for $\theta < \theta_c$--see below). The presence of a non-vanishing dimer correlation in \eqref{result1} implies spin-Peierls-like physics occurs simultaneously with Neel order. This Neel + dimer phase is a consequence of the completely broken spin rotational symmetry realized by a uniform Dzyaloshinkii-Moriya interaction, and uniform plus staggered magnetic fields. Schematically, the obtained phase (\ref{result1}) is depicted in the Fig. \ref{fig:kagome}.

The same phase, namely Neel + dimer, was obtained in double-frequency sine-Gordon model,\cite{Fabrizio:npb00} and in the work by Garate and Affleck on spin chains with uniform Dzyaloshinkii-Moriya interaction, uniform Zeeman magnetic field and exchange anisotropy,\cite{Garate} both strictly one-dimensional systems. In particular, the Neel+dimer phase reported in Ref.[\onlinecite{Garate}] possess Neel order which is perpendicular to the applied field (in our case played by the exchange field of the middle spins) and the DM direction.  These are the same features we find in our \eqref{result1}.  In particular, the small $q$ limit of our \eqref{H_odd_rot} is rather similar to Eq.(2.2) of Ref.[\onlinecite{Garate}].  The chief physical differences is that our case has staggered magnetic field and staggered DM interaction acting along the chain, while the Hamiltonian in Ref.[\onlinecite{Garate}] considers a uniform field and DM interaction.

\begin{figure}
\includegraphics[width=0.9\linewidth]{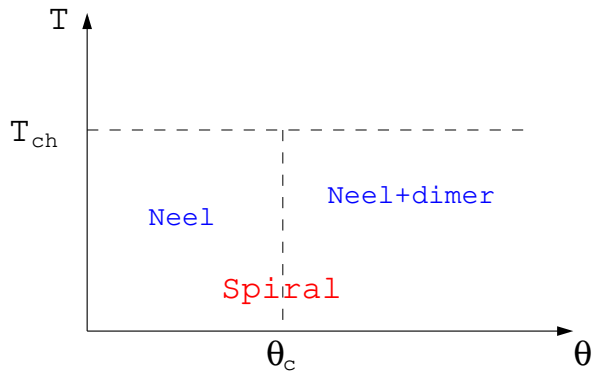}
\caption{(color online) Schematic phase diagram below the spin-chain ordering temperature, $T_{ch}$, as a function of the Dzyaloshinskii-Moriya interaction parameterized by the angle $\theta$, as given in \eqref{eq:theta} with $\chi_c=3\theta_c$.  ``Spiral" refers to the order of the middle spins which is established at a temperature scale of $T_m\sim (J')^2/J\gg T_{ch}$, while ``Neel" and ``Neel+dimer" refers to the dominant order (on top of the spiral) along the spin chains below a temperature $T_{ch}\sim (J')^4/J^3$.   The ``Neel" phase is described by \eqref{result3}, while the ``Neel + dimer" phase is described by  \eqref{result1}. }\label{fig:phase_diagram}
\end{figure}

In the opposite physical limit, when the coupling constant $\gamma_{x}$ dominates for $\theta <\theta_c$, we find that the order is given by
\begin{eqnarray}\label{result3}
\langle S^{x}_{x,y}\rangle &=& -\frac{h^{z}}{2\pi u} \cos(\phi) \cos(qx + (-1)^{x+y}\theta),  \nonumber\\
\langle S^{y}_{x,y}\rangle &=& -\frac{h^{z}}{2\pi u} \cos(\phi) \sin(qx + (-1)^{x+y}\theta), \nonumber\\
\langle S^{z}_{x,y}\rangle &=& -(-1)^{x+y}M_{2},
\end{eqnarray}
where $M_{2} \sim \left(\frac{J^{\prime}}{J}\right)^{2}$. In addition to the predominant spiral order in the $xy$-plane, this phase is characterized by a Neel order along the $z$-direction with moment $M_2$. The result \eqref{result3} is consistent with the prediction of Ref.~[\onlinecite{Oleg4}], which does not include the Dzyaloshinskii-Moriya interaction (vanishes in the limit $\theta \to 0$).  These results are summarized in Fig.\ref{fig:phase_diagram}.  The main role of the Dzyaloshinskii-Moriya interaction in the Neel phase is to choose the direction of Neel order parameter.   However, it also enters the spiral configuration (\ref{result3}) of spin chains via the argument of the $sin$ and $cos$.

\section{Conclusions}
\label{sec:conclusions}

In summary, we have studied the spatially anisotropic spin-1/2 kagome antiferromagnet with a staggered Dzyaloshinskii-Moriya (DM) interaction as shown in Fig.\ref{fig:kagome}.   As a function of increasing DM interaction, there is a phase transition from a spiral phase in the plane of the kagome lattice with an out-of-plane Neel order to a phase with with spiral order and an in-plane Neel order coexisting with dimer order.  Our results are summarized in Fig.~\ref{fig:phase_diagram}.  A rough numerical estimate the the DM interaction need to enter the ``Neel+dimer" phase suggest that it may be in an experimentally accessible range ($J'/J\approx 0.7$), and could possibly be tuned through the application of pressure to a spatially anisotropic system.  We hope this work will help inspire further experimental studies of spatially anisotropic kagome lattices.

It might also be interesting to realize a phase where both Neel (out-of-plane) and spin-Peierls orders coexist for an anisotropic two-dimensional lattice. It would seem that necessary conditions for such phase is a fully broken $SU(2)$ spin rotational symmetry (possibly  obtained by the application of an external magnetic field) as well as spin anisotropy-inducing terms such as DM interaction.  We leave that possibility as a topic for future study.

\acknowledgements
We are grateful to Victor Chua, Suhas Gangadharaiah, Philippe Lecheminant, and Oleg Starykh for discussions, and ARO grant W911NF-09-1-0527 and NSF grant DMR-0955778 for financial support.

\appendix
\section{Non-Abelian bosonization and Fusion rules}
\label{app:fusion}
The operators  used in the paper are defined by
\begin{eqnarray}
& {\bf J}_{R} = \frac{1}{2} R^{\dag}_{\alpha}{\bm \sigma}_{\alpha,\beta} R_{\beta}, \nonumber\\
& {\bf J}_{L} = \frac{1}{2} L^{\dag}_{\alpha}{\bm \sigma}_{\alpha,\beta} L_{\beta}, \nonumber\\
& {\bf M} = {\bf J}_{R} + {\bf J}_{L}, \nonumber\\
& {\bf N} = \frac{1}{2}R^{\dag}_{\alpha}{\bm \sigma}_{\alpha,\beta} L_{\beta} + h.c. ,\nonumber\\
& \epsilon = \frac{i}{2}(R^{\dag}_{\alpha}L_{\alpha} - h.c.),
\end{eqnarray}
where $R$ and $L$ denote fields of right- and left-moving fermions in one dimension.  See Ref.[\onlinecite{Gogolin}] for their definition in terms of bosonic fields. All of the operators defined above are functions of position and time $(x, \tau)$, which we have not explicitly indicated. Correlation functions of these fields are given by the following expressions
\begin{eqnarray}
& F_{R} = -\langle TR_{\alpha}(x,\tau)R^{\dag}_{\alpha}(0,0) \rangle = -\frac{1}{2\pi u (\tau-ix/u + \alpha\sigma_{\tau})} \nonumber\\
& F_{L} = -\langle TL_{\alpha}(x,\tau)L^{\dag}_{\alpha}(0,0) \rangle = -\frac{1}{2\pi u (\tau+ix/u + \alpha\sigma_{\tau})} \nonumber
\end{eqnarray}
where $\alpha = a_{0}/u$ is short time cut-off,  $a_{0}$ is a short distance cut-off  (distance between neighboring spins on a lattice), $\sigma_\tau={\rm sgn}(\tau)$, and $u=\pi Ja_0/2$. Below we
set $y =(x, \tau)$ for further convenience. The operators obey the fusion rules
\begin{eqnarray}
& J^{a}_{R/L}(y_{1})J^{b}_{R/L}(y_{2}) = -F_{R/L}(y)i\epsilon^{abc}J^{c}_{R/L}(Y) + \frac{1}{2}\delta^{ab} F^{2}_{R/L}(y), \nonumber\\
& J^{a}_{R/L}(y_{1})N^{b}(y_{2}) = -\frac{i}{2}F_{R/L}(y)[\epsilon^{abc}N^{c}(Y) \mp \delta^{ab} \epsilon(Y)], \nonumber\\
& J^{a}_{R/L}(y_{1})\epsilon(y_{2}) = \mp \frac{i}{2}F_{R/L}(y)N^{a}(Y),
\end{eqnarray}
from which the one-loop renormalization group flows are obtained.  Above, $y=y_{1}-y_{2}$ and $Y=(y_{1}+y_{2})/2$.

\section{Derivation of renormalization group equations}
\label{app:RG_second}
In this appendix we show one example of a derivation of the second order renormalization group (RG) equations of a relevant inter-chain interaction, such as those given in Eq.\eqref{eq:RG}.  To obtain the RG equations one has to plug (\ref{interchainrelevant_gen}) in to (\ref{eq:partition}).  As an example we focus on the second order term
\begin{widetext}
\begin{eqnarray}
2\frac{1}{2}\gamma_{M1}\gamma_{\partial N1}\int dx_{1}dx_{2}d\tau_{1}d\tau_{2}~
M_{y}^{z}(y_{1})M_{y+1}^{z}(y_{1})\partial_{x_{2}}N_{y}^{z}(y_{2})\partial_{x_{2}}N_{y+1}^{z}(y_{2}),
\end{eqnarray}
where, as in the previous Appendix, the coordinates $y_{1(2)}$ stand for $y_{1,2}=(x_{1},\tau_{1})$. Using the fusion rules of current operators one gets
\begin{eqnarray}
M_{y}^{z}(y_{1}) \partial_{x_{2}} N_{y}^{z}(y_{2}) = \frac{i}{2}\epsilon(X,t) \partial_{x_{2}}\left(F_{R}(y_{1}-y_{2}) - F_{L}(y_{1}-y_{2}) \right)
=\frac{1}{4\pi}\epsilon(X,t)\left(\frac{1}{(x+iu\tau)^2} + \frac{1}{(x-iu\tau)^2} \right).
\end{eqnarray}
The resulting integral is then
\begin{eqnarray}
&\gamma_{M1}\gamma_{\partial N1}\frac{1}{(4\pi)^2}\int dX dt~ \epsilon_{y}(X,t)\epsilon_{y+1}(X,t)\int_{-\infty}^{\infty} dx \int d\tau~ \left(\frac{1}{(x+iu\tau)^{2}} + \frac{1}{(x-iu\tau)^{2}} \right)^{2}  \nonumber\\
&= \gamma_{M1}\gamma_{\partial N1}\frac{d\ell}{8\pi u^{3}\alpha^{2}}\int dX dt~ \epsilon_{y}(X,t)\epsilon_{y+1}(X,t),
\end{eqnarray}
where $X=(x_{1}+x_{2})/2$, $x=x_{1}-x_{2}$, $t=(\tau_{1}+\tau_{2})/2$, and $\tau = \tau_{1}-\tau_{2}$. The integration is over $\tau$ is $\alpha < |\tau| < b\alpha$. One has this term in the renormalization group equation for $\gamma_{\epsilon}$ in the third line of (\ref{eq:RG}).
\end{widetext}


%

\end{document}